\documentstyle[floats,aps,graphicx,twocolumn,fancyheadings]{revtex}

\newcommand{\ud}{\,{\mathrm d}}
\newcommand{\zc}{\overline{z}}
\newcommand{\wc}{\overline{w}}
\newcommand{\fc}{\overline{f}}

\begin{document}
\draft
\title{Two-dimensional topological solitons in small exchange-dominated 
  cylindrical ferromagnetic particles}
\author{Konstantin L. Metlov}
\address{Institute of Physics ASCR, Na Slovance 2, Prague 8, CZ-18221}
\date{\today}
\maketitle
 
\begin{abstract}
  A general approach allowing to construct the magnetization
  distributions of arbitrary topological charge in small
  exchange-dominated cylindrical ferromagnetic particles is presented.
  The exchange energy functional is minimized by these distributions
  exactly. The magnetostatic energy is accounted partially, so that it
  facilitates a choice between the topologically equivalent
  exchange-only solutions.  The resulting magnetization distributions
  can be easily generalized to a variety of non-circular cylindrical
  shapes by means of a conformal transformation. As an example a
  magnetic structures of a thin circular ferromagnetic cylinder both
  with centered and displaced magnetic vortex and of a finite Bloch
  domain wall in an elongated particle is given.
\end{abstract}

\noindent {\it PACS \#\ \   75.60.Ch, 75.70.Kw, 85.70.Kh}


The magnetic structures of small ferromagnetic elements recently
attained great experimental interest due to their possible
applications in the integrated semiconductor-magnetic circuits and the
magnetic random access memory (MRAM). From the point of view of these
applications particularly the thin film elements (e.g. particles
having cylindrical shape) are more interesting than others because
they are more closely compatible with the current production
technology. Currently, a lot of attention was devoted to modelling the
magnetic structure of these
elements\cite{IS89_cyl,A90_2,UP92,GD94,CR97} (see also references
therein) with two main systematic approaches: the usage of
guessed trial functions and the numerical solution of the
Landau-Lifshitz equations. In this work an approximate (in accounting
for magnetostatic interaction) but analytic approach for solving this
task for a variety of element shapes is presented.

Mathematically, the problem of finding the static distribution of the
magnetization vector $\vec{M}(\vec{R})$ ($\vec{R}=\{X,Y,Z\}$ denotes
the radius vector of a point in space) within an isotropic
ferromagnetic particle in zero external magnetic field can be
formulated as finding the minimum of the following energy functional
of $\vec{m}(\vec{R})=\vec{M}(\vec{R})/M_S$:
\begin{equation}
  \label{eq:energy}
  M_S^2 e[\vec{m}] =  
    \int_V \left\{ \frac{L_E^2}{2} \sum_{i=X,Y,Z} (\vec{\nabla} m_i)^2 - 
  \vec{h}_D[\vec{m}]\cdot\vec{m} \right\} \ud^3 \vec{R},
\end{equation}
with the constraint $|\vec{m}(\vec{R})|=1$ for $\vec{R} \in V$, where
$V$ denotes the volume of the particle, $L_E=\sqrt{C/M_S^2}$ is the
exchange length, $C$ is the exchange constant of the material, $M_S$
is the saturation magnetization of the material,
$\vec{\nabla}=\{\partial/\partial X,\partial/\partial
Y,\partial/\partial Z\}$ is the gradient operator,
$\vec{h_D}[\vec{m}]=\vec{H}_D[\vec{m}]/M_S$ is the demagnetizing field
(a functional of $\vec{m}$) created by the magnetization distribution.
The demagnetizing field can be expressed with the help of Maxwell
equations through its scalar potential
$\vec{H}_D=\vec{\nabla}U(\vec{R})$, which is, in turn, a solution of
the Poisson equation $\vec{\nabla}^2 U = \vec{\nabla}\cdot\vec{M}$
with the requirement (due to the finite size of the particle) that
both $|\vec{R}| U$ and $|\vec{R}|^2 |\nabla U|$ are finite as
$|\vec{R}| \rightarrow \infty$, \cite{Aharoni_book}. The main
difficulty in treating this problem analytically arises from the fact
that the dipolar interaction is non-local (as expressed by a
functional dependence of $\vec{h_D}$ on $\vec{m}$), and the
Euler-Lagrange condition for the extremum of the functional
(\ref{eq:energy}) yelds a system of non-linear three-dimensional integral
partial differential equations.

The simplified variant of the problem when the demagnetizing field is
assumed to be zero $\vec{h_D} \equiv 0$ (so that the distribution is
defined by a system of partial differential equations, but still
non-linear ones) and $V$ is taken as infinite two-dimensional space
with the constraint $\vec{m}(\vec{R})=m_0$ at $|\vec{R}|\rightarrow
\infty$ was solved by Belavin and Polyakov \cite{BP75,KIK90}. Let us
briefly rederive this solution because the procedure will be relevant
further in this paper. Noting that the
constraint $|\vec{m}|=1$ leaves only two independent components of the
magnetization vector, and parametrizing the magnetization vector by a
complex function $w(z,\zc)$ of a complex variable $z=X+\imath Y$,
$\zc=X-\imath Y$, $\imath=\sqrt{-1}$ so that $m_x+\imath m_y = 2
w/(1+w\wc)$ and $m_z=(1-w\wc)/(1+w\wc)$, the exchange energy functional
can be written as
\begin{equation}
  \label{eq:energy_complex}
  e  =  L_E^2 L_Z \int_V \frac{4}{(1+w\overline{w})^2}
  \left(
    \frac{\partial w}{\partial z}
    \frac{\partial \overline{w}}{\partial\overline{z}}+
    \frac{\partial w}{\partial \overline{z}}
    \frac{\partial \overline{w}}{\partial z}
  \right) \ud^2 \vec{R_2},
\end{equation}
where $\partial/\partial z=(\partial/\partial X - \imath
\partial/\partial Y)/2$, $\partial/\partial
\overline{z}=(\partial/\partial X + \imath \partial/\partial Y)/2$ and
$\vec{R_2}$ is a two-dimensional radius vector. The system of Euler
equations for the extremum of this functional noting that the
variables $z$ and $\zc$ are independent is
\begin{equation}
  \label{eq:energy_euler}
  \frac{\partial}{\partial z}
  \left(\frac{\partial w}{\partial \overline{z}}\right) =
  \frac{2 \overline w}{1+w\overline{w}}     
  \frac{\partial w}{\partial z}
  \frac{\partial w}{\partial\overline{z}}.
\end{equation}
It is immediately clear that any analytic function $w$ of the complex
variable $z$ satisfies this equation. Because the Cauchy-Riemann
conditions for the analyticity $\partial w/\partial\zc = 0$, if
satisfied, turn both sides of (\ref{eq:energy_euler}) to zero.  The
Belavin-Polyakov (BP) soliton corresponds to the representation of
$w(z)$ as a rational function of $z$. Zeroes of $w(z)$ correspond to
the centers of solitons (at these points $m_z=1$), at the infinity
$|z|\rightarrow \infty$ the BP soliton has $m_z=-1$.

Let us now turn to small particles in the form of cylinders,
choosing the coordinate system in such a way that the axis $Z$ is
perpendicular to the faces of the cylinder.  As long as the thickness
of these cylinders is much smaller than the exchange length it is
reasonable to assume that the magnetization distribution is uniform
along the thickness ($Z$ axis). This assumption immediately reduces
the dimensionality of the problem and allows us to use the
$\vec{m}(\vec{r})$ parametrization in terms of $w(z,\zc)$.

\begin{figure}[htbp]
  \begin{center}
    \includegraphics[width=7.0cm]{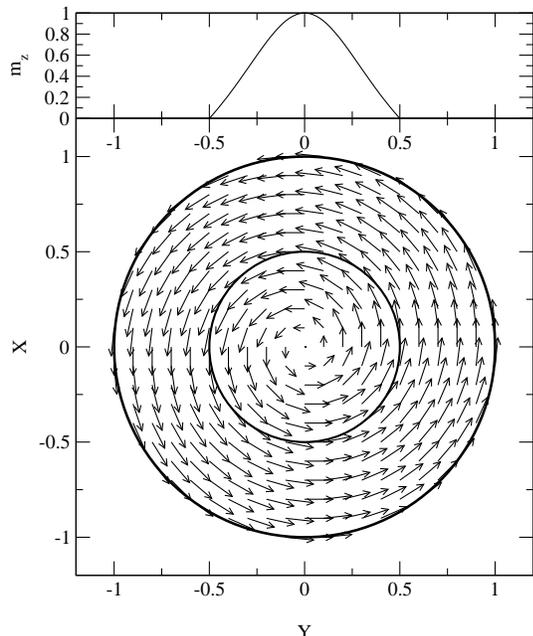}
    \caption{The structure of the vortex with $\nu=1$ and $c=1/2$ centered
      in the cylindrical particle. The outer solid circle is the
      contour of the particle, the inner circle is the boundary of the
      vortex core.}
  \end{center}
  \label{fig:vortex_centered}
\end{figure}

A special case of the problem (\ref{eq:energy}) taking into account
the demagetizing field was solved by Usov \cite{UP93} for the particle
shaped as a circular cylinder and is also recently confirmed
experimentally \cite{SOHSO00}. This solution has the topology of a vortex,
see Fig.~\ref{fig:vortex_centered}. Its interesting property is that
only a part of the magnetization distribution is a BP soliton (Usov
calls this part the vortex core), the other part (represented in
cylindrical coordinate system) as $m_r=1$ is not. The component of
magnetization normal to the cylinder face $m_z$ changes from $m_z=1$
at the center of the magnetic vortex to $m_z=0$ at the vortex core
boundary, and continues to stay zero in the non-core region.

The non-core part of the solution by Usov can be easily understood on
the basis of the magnetic pole avoidance principle
\cite{Aharoni_book}, which is due to the fact that the energy of the
demagnetizing field $\vec{h_D}$ is always positive, and, since this
field is created by magnetic poles (whose density is
$\vec{\nabla}\cdot\vec{M}$), the system tries to avoid having poles.
At a boundary of a finite magnetic particle where $|m|$ abruptly
changes from $1$ to $0$ (no magnetic material) the density of magnetic
poles is proportional to the component of the magnetization vector
normal to the boundary.  Thus, compared to the BP solution in the
whole circular cylinder the solution by Usov has less magnetic poles
on cylinder faces and is more favorable from the point of view of
magnetostatic energy.  Unfortunately, the procedure employed by Usov
can hardly be generalized to other particle shapes because it heavily
relies on the cylindrical coordinate system.

If we look back at equation (\ref{eq:energy_euler}) we can easily
suspect that it may have more solutions besides the BP solitons.
Although it was shown that the BP solitons are the only solutions with
the finite total exchange energy in the infinite space, this
requirement becomes irrelevant as long as we consider finite
particles. Another family of solutions of (\ref{eq:energy_euler})
was first found by Gross \cite{G78} and is called merons: they are
expressed as $w_M(z,\zc)=f(z)/\sqrt{f(z)\fc(\zc)}$, where $f(z)$ is
again an arbitrary analytic function. These solutions have $m_z=0$
everywhere and do not produce magnetic poles on the cylinder
faces.  However, they diverge at the points where $f(z)=0$ so that the
exchange energy density is infinite at these points, therefore, the
pure meron solutions can not be realized in the particles.

We may note, however, that on the contour $|f(z)|=1$ the meron
continuously joins the BP soliton $w(z)=f(z)$. Thus, the function
\begin{equation}
  \label{eq:sol_SM}
  w_{SM}(z,\overline{z})=\left\{
    \begin{array}{ll}
      f(z) & |f(z)| \leq 1 \\
      f(z)/\sqrt{f(z) \fc(\zc)} & |f(z)|>1
    \end{array}
    \right. 
\end{equation}
is also a solution of (\ref{eq:energy_euler}) for any analytical
function of complex variable $f(z)$, is regular and continuous
everywhere and has much less magnetic poles then the BP solitons. This solution for the magnetization distributions of
small magnetic particles is the main result of this paper. By choosing
the appropriate $f(z)$ many realistic magnetization distributions can
be obtained, some of the examples will be given later in the text.

The solution (\ref{eq:sol_SM}), generally, should have lower
magnetostatic energy then the corresponding BP soliton $w(z)=f(z)$
because of less magnetic charges on the faces of the cylinder.
However, such an advantage may turn out as unimportant for cylinders
with a very small area of the region $f(z)>1$. Unfortunately, the
magnetostatic energy is strongly shape dependent due to its long-range
character and the estimates of this critical particle size need to be
done separately in each case. Once the energy of the solution
(\ref{eq:sol_SM}) is lower, the BP soliton is unstable with respect to
the continuous transformation to (\ref{eq:sol_SM}) by tilting spins in
the region $|f(z)|>1$ towards the plane of the cylinder. Such a
transformation does not have an associated energy barrier because
$w_{SM}(z)$ and $f(z)$ have the same topology.

Another important fact, easily checked by direct substitution, is that
for any analytic function $o(z)$, the function
$w(o(z),\overline{o}(\zc))$ solves the equation
(\ref{eq:energy_euler}) if $w(z,\zc)$ solves it. This property allows
us to generalize the solution (\ref{eq:sol_SM}) using conformal
transformations to particle shapes where the function $f(z)$ is hard
to choose on the basis of symmetry argument alone.

As an illustration consider a particle shaped as a circular cylinder.
From the symmetry it is natural to choose the BP solution in the form
of a vortex centered in the cylinder (chosen as the origin of the
coordinate system), which is $f(z)=\imath (z/c)^{|\nu|}$, where $c$ is
an arbitrary real number defining the scale and $\nu$ is the integer
topological charge of the vortex. In the case $\nu=1$ this solution is
equivalent to the one obtained and investigated both analytically and
numerically by Usov\cite{UP93,UP94} and shown in
Fig.~\ref{fig:vortex_centered}. The definite value of $c$ for a
particular material and the geometry of the particle can be found by
minimizing the total energy (\ref{eq:energy}) with respect to $c$
similarly as it is done in \cite{UP93,UP94}.

\begin{figure}[htbp]
  \begin{center}
    \includegraphics[width=7.0cm]{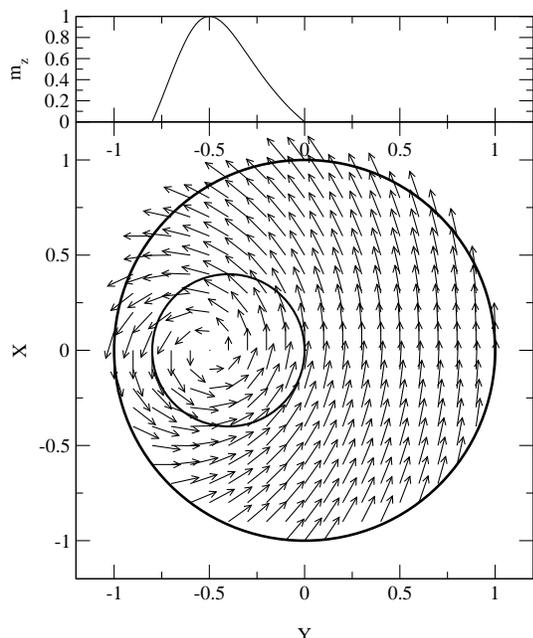}
    \caption{The structure of the vortex with $\nu=1$ and $c=1/2$ displaced
      by 1/2 with respect to the particle center. Note that the vortex
      core has the circular boundary but is otherwise asymmetrical.}
  \end{center}
  \label{fig:vortex_displaced}
\end{figure}

To displace the vortex the conformal transformation of the unit circle
to the unit circle can be used, which is expressed by the rational
function $o(z)=(z-a)/(1-\overline{a} z)$, where $a=X_c+i Y_c$ is the
new coordinate of the vortex center, $|a|<1$. The resulting
configuration (\ref{eq:sol_SM}) with $f(z)=\imath
(z-a)/(1-\overline{a} z)/c$ is shown in
Fig.~\ref{fig:vortex_displaced}. It can be seen from the picture that
this configuration corresponds well (at least qualitatively) to the
application of an external field in the direction of $Y$ axis. Also
the case of $|a| \rightarrow 1$ corresponds to the case of the
uniformly magnetized particle. The complete analysis of the energies
(both Zeeman and the magnetostatic) and the stability of such a
displaced vortex in an external magnetic field will be published
elsewhere\cite{GM01}.

Another example is a multi-vortex solution of an arbitrary topological
charge, which can be built by choosing $f(z)$ in (\ref{eq:sol_SM}) as
a product of powers $(z-z_i)^{\nu_i}$, where $z_i$ and $\nu_i$ are the
coordinate of the center and the topological charge of $i$-th vortex,
respectively. In particular, when there is an infinite number of
$\nu_i=1$ vortices sitting on a straignt line (say $0<z<1$) the
resulting magnetization distribution $f(z)=\imath \exp(\int_0^1
\log(z-x) \ud x)=(z-1)(z/(z-1))^z/e$ can be classified as a finite
Bloch domain wall with ends at points $0$ and $1$. The curves
tangential to the vector $\vec{m}$ are of elongated (but not
exactly elliptical) shape. A similar distribution was indeed
observed in elongated magnetic dots \cite{WRNS00} and interpreted as a
two-domain configuration.

To summarize, a general approach allowing to build the magnetization
distributions with an arbitrary topological charge in small
exchange-dominated cylindrical magnetic particles of variety of shapes
is presented. The produced magnetization distributions minimize the
exchange functional exactly and partially account for the
magnetostatic interaction. There are two factors limiting the
presented consideration to small particles only. The first is
connected with the fact that the magnetization distribution is assumed
to be uniform along the cylinder thickness.  The second is that the
exchange energy of the proposed solution (\ref{eq:sol_SM}) within a
circle of a given radius increases logarithmically with the radius. It
means that the exchange energy diverges with increasing the
particle size, thus other non-uniform configurations may become more
energetically favorable in larger particles.

The presented approach should coexist well with existing numerical
finite-element methods of micromagnetics as a way to specify the
initial configuration with given topological properties, and to
subject it to a final numerical refinement.

This work was supported in part by the Grant Agency of the Czech
Republic under projects 202/99/P052 and 101/99/1662. I would like to
thank Dr. Vladimir Kambersk{\'y}, Dr. Ivan Tom{\'a}{\v s}, and Dr. Konstantin
Guslienko for reading the manustcript and many valuable discussions.


\end{document}